-------------------------------------------------------------------------
\documentstyle[12pt]{article}
\newcommand{\be}{\begin{equation}}
\newcommand{\ee}{\end{equation}}
\newcommand{\bea}{\begin{eqnarray}}
\newcommand{\eea}{\end{eqnarray}}
\begin{document}
\begin{titlepage}
\begin{flushright}
DO-TH-93/15 \\
May 1993
\end{flushright}
 
\vspace{20mm}
\begin{center}
{\Large
Quantum Corrections to the Electroweak Sphaleron Transition}
\vspace{10mm}
 
{\large  J. Baacke and S. Junker} \\
\vspace{15mm}
 
{\large Institut f\"ur Physik, Universit\"at Dortmund} \\
{\large D-44221 Dortmund 50} \\ {\large Germany}
\vspace{40mm}
 
\bf{Abstract}
\end{center}
We have performed a new exact evaluation of the fluctuation
determinant $\kappa$ of the electroweak sphaleron with $\Theta _W =0$.
The results differ significantly from a previous calculation of this 
quantity by Carson et. al. . We find that $\kappa$ is of order
1 in units $(gv)^6$ while the previous results indicated a strong
suppression of the sphaleron transition by this factor.
\end{titlepage}
The electroweak theory has been found \cite{Man,Kli}
to possess a topologically nontrivial solution which describes
a saddlepoint between two topologically distinct vacua.
 
The recent interest in this solution has been centered around
its possible r\^ole in generating baryon number violating
processes in the early universe or even at accelerator energies
\cite{KuRuSha} - \cite{Ring}.
The rate of sphaleron transitions in the range of temperatures
$M_W(T) \ll   T \ll   M_W(T)/\alpha_W    $ has been 
derived by Arnold and McLerran \cite{ArnMcL}. It
is given by
\be
\Gamma=\frac{\omega_{-}}{2 \pi}{\cal N} e^{-E_{sph}/T}\kappa
\ee
Here $\omega_{-}$ is the absolute value of the eigenvalue of the
unstable mode, the prefactor ${\cal N} $ refers to normalizations
introduced by the translation and rotation zero modes. $E_{sph}$
is the sphaleron energy and the factor $\kappa$ takes account of
the quantum fluctuations of the sphaleron.
It is given by  
\be
\kappa = {\rm Im} (\frac{\Delta_{FP}^S \Delta_{gf}^0}{\Delta_{FP}^0
\Delta^{'S}_{gf}})^{1/2}  \label{kapdef} 
\ee
where the symbols $\Delta$ denote the small fluctuation
determinants of the gauge fixed action (gf) and the 
Fadeev-Popov action (FP) evaluated around the sphaleron (S)
and the vacuum ($0$), respectively. If the fluctuation operators 
are diagonalized the determinants are formally given by the
product of the squared eigenfrequencies $\lambda_{\alpha}^2$.
 The determinant $\Delta^{'S}_{gf}$  of the gauge fixed action 
is to be evaluated without the zero modes and with the 
eigenvalue of unstable mode, $\lambda_{-}^2=-\omega_{-}^2$ replaced
by its absolute value.
 
$\kappa$
 has been first evaluated  \cite{CarMcLe}
  using an approximation scheme developed by Diakonov et al.
  \cite{Dia}
and subsequently exactly  by Carson et al. \cite{CarLi} .
The results of this exact calculation differ significantly
from the approximate ones and from a perturbative estimate.
It is therefore of interest to repeat this exact evaluation
and this is the subject of our investigation. Technically
our evaluation differs essentially from the one in Ref. \cite{CarLi}
so that the calculations can be considered as independent.
While we use the background gauge as do  Carson et al.,
we use another angular momentum basis and a different
scheme for evaluating the determinant. In this short note we
will present only an outline of our approach and the numerical
results. A more extensive account is in preparation \cite{BaaJu}.
 
The analysis of the small fluctuations around the sphaleron
requires a partial wave decomposition with respect to the
quantum number $\vec{K} = \vec{J} + \vec{I}$ ($K$-spin).
Our present work is based on the analysis of Ref. \cite{BaaLa}
where the sphaleron stability was investigated.
The small fluctuation equations obtained there had to be modified
however. While that investigation was performed in
the $A_0=0$ gauge we use here the background gauge. So
a gauge fixing term had to be added and also we had to
construct the small fluctuation Lagrangean for the Fadeev-Popov
modes. The basic algebra and the resulting equations will be presented
elsewhere \cite{BaaJu}.

The evaluation of the determinant has been performed using the
Euclidean Green function in analogy to some recent investigations
of one of the authors (J.B.) \cite{BaaEff}. We will sketch the method
for the case of a single field:
The logarithm of the fluctuation determinant around the vacuum,
divided by the one around the sphaleron can be written
formally (postponing
renormalization and the treatment of zero and unstable modes) as
\be
\ln ((\frac{\Delta^0}{\Delta^S})^{1/2})=
\int_0^{\infty}d\nu \nu \sum_{\alpha}(\frac{1}
{\nu^2+(\lambda^S_{\alpha})^2}-\frac{1}{\nu^2+(\lambda^0_{\alpha})^2})
\label{bas1}
\ee
where $\lambda_{\alpha}^{S,0}$ are the eigenfrequencies of the
fluctuations around sphaleron and vacuum, respectively.
Further we can can relate the integrand to Euclidean Green functions
via
\be
\sum_{\alpha}\frac{1}{\nu^2+\lambda_{\alpha}^2}= \nonumber \\
\int d^3x \sum_{\alpha}
\frac{\phi^*_{\alpha}(\vec x)\phi_{\alpha}(\vec x)}
{\nu^2+\lambda_{\alpha}^2}=  \\ \int d^3x G_E(\vec x,\vec x,\nu)
\ee
Here $\phi_{\alpha}$ are the eigenmodes and the Euclidean Green function
is a solution of the equation
\be 
(\nu^2 + {\cal D}^2)G_E(\vec x,\vec x',\nu )=\delta^3(\vec x -
\vec x')
\ee 
${\cal D}^2$ is the fluctuation operator $\delta^2 {\cal S}^3/
\delta\phi^2$ where $S^3$ is the three dimensional 
action and $\phi$ refers to the different field fluctuations
around vacuum and sphaleron.  
So the final equation for the fluctuation determinant is
\bea
\ln((\frac{\Delta^0}{\Delta^S})^{1/2}) & = &
\int_0^{\infty}d\nu \nu F(\nu) \nonumber \\
F(\nu)& = & \int d^3x (G^S_E(\vec x,\vec x,\nu)-
G^0_E(\vec x,\vec x,\nu))
\eea
 
The evaluation of the Green function can be done \cite{BaaEff}
by decomposing it into partial waves and by determining the
solutions $f^-$, regular at $r=0$,  and $f^+$, regular as
$r \rightarrow \infty$, of the corresponding differential equations.
The partial wave Green function is then obtained in the
usual way as
\be
G_l(r,r',\nu) = f_l^{-}(r_{<})f_l^{+}(r_{>})/[r^2 W(f_l^{-},f_l^{+})]
\ee
where $W$ is the Wronskian. Actually we have not calculated
the partial wave amplitudes $f_l^{\pm}$ but the amplitudes 
$h_l^{\pm}$ defined by
\be
 f_l^{\pm}(r)=[1+h_l^{\pm}(r)]b_l^{\pm}(\rho r)
\ee
Here $b_l^{\pm}$ are the solutions of the free partial wave equations
with the same boundary conditions as $f_l^{\pm}$, i. e. the
modified spherical Bessel functions $i_l$ and $k_l$, and
$\rho=(\nu^2+m^2)^{1/2}$.
Then we have
\be
G_l(r,r,\nu)-G_l^{(0)}(r,r,\nu)
  = \rho[h_l^+(r)+h_l^-(r)+h_l^+(r)h_l^-(r)]
    i_l(\rho r) k_l(\rho r)
\ee
The free part of the Green function,
instead of being subtracted, is omitted here from the outset,
avoiding the occurence of small differences of large contributions.
The detailed procedure for a multichannel system as we have it 
here is described extensively in Ref. \cite{BaaEff}.
 
Renormalization requires here just the
removal of the tadpole graphs  and replacing $M_W$ and $M_H$ 
by $M_W(T)$ and $M_W(T)$ as discussed in Ref. \cite{ArnMcL}.
If the tadpole contribution is subtracted from $F(\nu)$ the difference
 $F_{ren}(\nu)$ behaves
as $\rho^{-3}$ as $\nu \rightarrow \infty$ and the
$\nu$ integral is ultraviolet
convergent.

Of course $F_{ren}$ behaves as $ 6 \nu ^{-2}$ near zero
due to the presence of 6 zero modes, leading to a logarithmic
divergence of the $\nu$ integration. It can be realized that
Eq.(\ref{bas1}) has been constructed in such a way that
the contribution of each eigenmode is related to the evaluation of the
integral of $F_{ren}(\nu)$ at its lower limit; the procedure for
removing the zero mode contributions consists therefore in evaluating
the integral with a sufficiently small lower limit $\epsilon$
and then adding $6 \ln (\epsilon)$. The resulting expression
stays finite as $\epsilon \rightarrow 0$ and this limit is the final
answer:
\be
\ln \kappa = \lim_{\epsilon \rightarrow 0}
(\int_{\epsilon}^{\infty}d\nu \nu F_{ren}(\nu) + 6 \ln \epsilon).
\ee
It should be noted that $\nu$ and therefore $\epsilon$
have the dimension of energy, 
so the scale used for these variables matters. We have
performed our calculations in units of $M_W$ while in
Ref. \cite{CarLi} the scale was 
 $gv$ ; we had to take this into account for comparing our
results. In the complete rate the dimensional scale reappears
in the scale of the translation and rotation mode prefactors
so that the final result is unique as it should. 
 
It remains to consider the r\^ole of the unstable mode. It leads
to a pole $1/(\nu^2-\omega_{-}^2)$ in $F(\nu)$ where $\omega_{-}$
is the absolute value of the corresponding imaginary eigenvalue
$\lambda_{-}$. Since this eigenvalue should be included in the
determinant with its absolute value, the correct prescription 
for evaluating the integral is the principal value one. In praxi
we have subtracted from $F_{ren}(\nu)$ an expression
\be
F_{pole} =\frac{1}{\nu^2-\omega_{-}^2}-\frac{1}{\nu^2+\sigma^2}
\label{pole}
\ee
whith some suitable value for $\sigma$. This subtraction makes the
integrand  regular at $\nu = \omega_{-}$  without spoiling its
ultraviolet convergence. We have then added the analytic principal value
integral of this expression back to the numerical integral.
 
Our analytical and numerical procedure contains various
implicit checks:
 
\noindent 
(i) The expected asymptotic behaviour of $F(\nu)$
and $F_{ren}(\nu)$ which follow from analysing the perturbation
expansion of the Euclidean Green function. Note that
for $F(\nu)$ not only the
power behaviour but also its absolute normalization are
determined.

\noindent
(ii) The behaviour of the partial wave contributions at large K
which can be derived from a perturbative analysis to be
as $K^{-2}$.

\noindent
(iii) The occurence of the zero mode and unstable mode poles 
with the correct positions and residues.

Furthermore the differential equations for all partial waves
have been checked to a certain extent by verifying their gauge
invariance before gauge fixing \cite{BaaLa}. The decoupling
of the time components of the vector field after adding the
gauge fixing term represents a further check.
 
The numerical calculations were extended to angular momenta
(K spins) up to 25. The contributions of the higher partial waves
were included using an extrapolation $A/K^2+B/K^3$, where
$A$ and $B$ were obtained from a fit to the contributions
of $K=21$ to $K=25$. The $\nu$ integration was performed
numerically up to $\nu_{max} \simeq 2.5$, then an asymptotic part
was added to the integral by extrapolating $F_{ren}$ as
$C/\rho^3+D/\rho^5$.
 
Fig. 1 shows a typical plot for the functions $\nu F(\nu)$
and $ \nu F_{ren}(\nu)$ together with the expected asymptotic
behaviour at small and large $\nu$. The pole part $F_{pole}$
(see Eq. (\ref{pole})) has been subtracted in all amplitudes.
The results for $\ln\kappa$ are given in Table~I
for various values of $\xi = M_H/M_W$, whith $\kappa$ in units
of $M_W^6$.
They are plotted in Fig.~2 together with previous results
and estimates in the scale $gv$, i.e. after subtracting
$6\ln2$ from the values of Table~I. We think that the main
uncertainty of our results  comes from the extrapolation
of $F_{ren}(\nu)$ to $\nu = \infty$. We estimate the error
of this asymptotic contribution to be around $10 \%$ yielding
an typical error of $0.3$ for $\ln \kappa$.
 
Our results differ considerably from the ones of Carson et al.
\cite{CarLi} while for $M_H~\approx~M_W$ they are near the estimate
based on the method of Diakonov et al. \cite{Dia}. However
they show a different trend for small
Higgs masses. If they are correct - and in view of the mentioned
consistency checks we can have no reasonable doubt - this
means that the inclusion of the fluctuation determinant $\kappa$
does not suppress the sphaleron transition even for 
(unrealistic) small Higgs masses.
 
\section*{Acknowledgement}
 
It is a pleasure to thank Valeri Kiselev for useful discussions
in the final stage of this investigation and for reading the 
manuscript.

\vspace{20mm}
\begin{center} 
\section*{Table 1}
\vspace{6mm}
\begin{tabular}{|c||c|c|c|c|c|c|c|}\hline
$\xi$ & 0.4 & 0.5 & 0.6 & 0.8 & 1.0 & 1.5 & 2.0 \\  \hline
$\omega_-$ & 1.32 & 1.36 & 1.39 & 1.45 & 1.51 & 1.62 & 1.71 \\ \hline
$\ln \kappa$ & 6.18 & 5.89 & 5.68 & 5.50 & 5.48 & 5.62 & 5.71 \\ \hline
\end{tabular}
\end{center}
\newpage                                    

\section*{Figure Captions}

\vspace{6mm}
\noindent 
Fig. 1 {\bf The Function $F(\nu)$ for $\xi = M_H/M_W =1$}

The solid line shows $\nu F_{ren}(\nu)$, the dashed line
the unrenormalized $\nu F(\nu)$. The pole contribution
$F_{pole}$ of Eq.(\ref{pole}) has been removed. The dotted lines show
the expected power behaviours at small and large $\nu$ and
the dash-dotted line the analytically known
tadpole contribution to $\nu F(\nu)$.

\vspace{10mm}
\noindent
Fig. 2 {\bf The Fluctuation Determinant}

We plot the logarithm of the fluctuation determinant $\kappa$
for various values of the ratio $\xi = M_H/M_W$.
The circles are our results, the crosses those of
Carson et al.. The full line is the estimate of Carson and
McLerran based on the method of Diakonov et al. and the
dashed line a perturbative estimate.

\end{document}